\documentclass[twocolumn, prb, aps, showpacs, letterpaper]{revtex4}

\usepackage[dvips]{graphicx} 
\usepackage{subfigure}
\usepackage{latexsym}
\usepackage{dcolumn} 
\usepackage{bm} 
\usepackage{textcomp} 

\begin{document}
\title{
Transport and structural study of pressure induced magnetic states in Nd$_{0.55}$Sr$_{0.45}$MnO$_{3}$ and Nd$_{0.5}$Sr$_{0.5}$MnO$_{3}$
}

\author{Congwu Cui}
\author{Trevor A. Tyson}
\author{Zhiqiang Chen}
\affiliation{Physics Department, New Jersey Institute of Technology, Newark, New Jersey 07102
}
\author{Zhong Zhong}
\affiliation{National Synchrotron Light Source, Brookhaven National Laboratory, Upton, NY 11973
}

\date{\today}

\begin{abstract}
Pressure effects on the electron transport and structure of Nd$_{1-x}$Sr$_{x}$MnO$_{3}$ (x = 0.45, 0.5) were investigated in the range from ambient to $\sim$6 GPa. In Nd$_{0.55}$Sr$_{0.45}$MnO$_{3}$, the low temperature ferromagnetic metallic state is suppressed and a low temperature insulating state is induced by pressure. In Nd$_{0.5}$Sr$_{0.5}$MnO$_{3}$, the CE-type antiferromagnetic charge-ordering state is suppressed by pressure. Under pressure, both samples have a similar electron transport behavior although their ambient ground states are much different. It is surmised that pressure induces an A-type antiferromagnetic state at low temperature in both compounds.
\end{abstract}

\pacs{75.47.Lx, 62.50.+p, 71.27.+a, 75.25.+z}

\maketitle

\section{Introduction}
In Nd$_{1-x}$Sr$_{x}$MnO$_{3}$ manganite, the size difference between Nd$^{3+}$ and Sr$^{2+}$ is large ($\sim$0.15 \AA). With increasing Sr$^{2+}$ concentration, the bandwidth increases. With changes in x, intriguing spin, charge, and orbital phases are produced and extensive studies have been performed on these systems.\cite{tokura_jmmm_200_1_99}

In the x = 0.5 compound, on cooling from room temperature, a transition from a paramagnetic insulating (PMI) phase to a ferromagnetic metallic (FMM) phase occurs at $\sim$255 K and a transition from FMM phase to charge-ordered (CO) antiferromagnetic insulating (AFI) phase is observed at $\sim$155 K. The magnetic structure in the CO AFI phase is CE-type.\cite{kawano_prl_78_4253_97} With the application of a magnetic field, the FMM state is enhanced and the charge-ordering state is suppressed completely above 7 T.\cite{kuwahara_sci_270_961_95} The magnetic field induced collapse of CO state is accompanied by a structural transition in which the volume increases drastically, leading to large positive magnetovolume effect.\cite{mahendiran_prl_82_2191_99} Orbital ordering coincides with charge-ordering. Different orbital ordering types, $d_{3x^{2}-r^{2}}/d_{3y^{2}-r^{2}}$-type\cite{nakamura_prb_60_2425_99} or $d_{x^{2}-y^{2}}$-type,\cite{zayagin_solidstate_121_117_02} have been suggested in this compound.

Nd$_{0.45}$Sr$_{0.55}$MnO$_{3}$ is an A-type antiferromagnetic metal with coupled magnetic and structural transition at $\sim$225 K.\cite{kawano_prl_78_4253_97} The Mn moments are ferromagnetically aligned in the ab-plane in Pbnm symmetry. Charge carriers are confined within the ab-plane while the transport along the c-axis is quenched, leading to highly anisotropic resistivity ($\rho_{c}/\rho_{ab}$$\sim$10$^{4}$ at 35 mK).\cite{kuwahara_mrssp_494_83_98} The antiferromagnetic spin ordering is accompanied by the $d_{x^{2}-y^{2}}$-type orbital ordering, both of which are simultaneously destroyed by high magnetic fields, concomitant with a discontinuous decrease of resistivity.\cite{hayashi_prb_65_024408_02}

In this manganite system, the magnetic, electronic and orbital transitions are correlated with an abrupt structural transition, in which the \textit{a} and \textit{b} lattice parameters are elongated and the \text{c} parameter is compressed (in Pbnm symmetry).\cite{kuwahara_mrssp_494_83_98} Other groups showed that in Nd$_{0.5}$Sr$_{0.5}$MnO$_{3}$, during the transition from FMM to AFI CO state, the crystal symmetry is lowered to monoclinic P21/m.\cite{eremenko_ltp_27_930_01,laffez_mrb_31_905_96} Ritter \textit{et al}.\cite{ritter_prb_61_9229_00} suggested that the low temperature AFI CO phase is phase-segregated into two different crystallographic structures and three magnetic phases: orthorhombic (Imma) ferromagnetic, orthorhombic (Imma) A-type antiferromagnetic, and monoclinic (P21/m) charge-ordered CE-type antiferromagnetic phases, in which a magnetic field can induce the charge-ordered monoclinic phase to collapse and to transform into the FMM orthorhombic phase. Kajimoto \textit{et al}.\cite{kajimoto_prb_60_9506_99} showed that in the CE-type and A-type antiferromagnetic states, the MnO$_{6}$ octahedra are apically compressed corresponding to $d_{3x^{2}-r^{2}}/d_{3y^{2}-r^{2}}$ or $d_{x^{2}-y^{2}}$ orbital ordering.

Hydrostatic and uniaxial pressures affect the CO and FMM states differently. In Nd$_{0.5}$Sr$_{0.5}$MnO$_{3}$, hydrostatic pressure ($<$$\sim$1 GPa) increases T$_{C}$ with dT$_{C}$/dP = 6.8 K/GPa and decreases T$_{CO}$ at a rate of 8.4 K/GPa,\cite{moritomo_jpsj_66_556_97} while uniaxial pressure along the c-axis decreases T$_{C}$ at a rate of 60 K/GPa and increases T$_{CO}$ at a rate of 190 K/GPa.\cite{arima_prb_60_15013_99} In Nd$_{0.45}$Sr$_{0.55}$MnO$_{3}$, with the application of uniaxial pressure along the c-axis, the A-type antiferromagnetic phase is stabilized by increasing T$_{N}$ at 66 K/GPa, implying the stabilization of the $d_{x^{2}-y^{2}}$ orbital.\cite{arima_prb_60_15013_99} In thin films, due to the effect of substrate, biaxial strain can be induced. In Nd$_{0.5}$Sr$_{0.5}$MnO$_{3}$ thin films, the thickness dependent strain tunes the competition between CO insulating and FMM states,\cite{prellier_apl_75_397_99} and there are optimal strain conditions under which the CO or metallic states appear.\cite{qian_prb_63_224424_01}

The correlation between the structure and the electronic and magnetic transitions indicates its crucial role in this doping system. We have studied the CO, FMM, and antiferromagnetic state changes in Nd$_{0.55}$Sr$_{0.45}$MnO$_{3}$ and Nd$_{0.5}$Sr$_{0.5}$MnO$_{3}$ through high-pressure (up to $\sim$6 GPa) resistivity and structure measurements. It is found that pressure induces similar electronic and magnetic behavior in them, which can be partially related to the changes of orthorhombic distortion under pressure.

\section{Samples and Experimental Methods}

The samples were prepared by solid-state reaction. Stoichiometric amounts of Nd$_{2}$O$_{3}$, MnO$_{2}$, and SrCO$_{3}$ powder were mixed, ground, and calcined at 900 \textcelsius\space for 15 hours. The sample was then cooled down to room temperature and reground and then calcined again at 1200 \textcelsius\space for 17 hours. The powder was then pressed into pellets. The pellets were sintered at 1500 \textcelsius\space for 12 hours, cooled down to 800 \textcelsius\space at a rate of 5 \textcelsius/min, and then quickly cooled down to room temperature. The pellets were annealed at 1200 \textcelsius\space and cooled down slowly to room temperature at 1 \textcelsius/min. The x-ray powder diffraction patterns show a single crystallographic phase for each sample. The magnetization and resistivity measurements are consistent with the results of other groups.\cite{tokura_jmmm_200_1_99, caignaert_solidstate_99_173_96, tomioka_prl_70_3609_97} The details of the high-pressure resistivity and high-pressure x-ray diffraction measurement methods and error analysis were described previously.\cite{cui_prb_67_104107_03}

\section{Results and Discussions}

\subsection{Nd$_{0.55}$Sr$_{0.45}$MnO$_{3}$}
\begin{figure}
\includegraphics[width=2.2in]{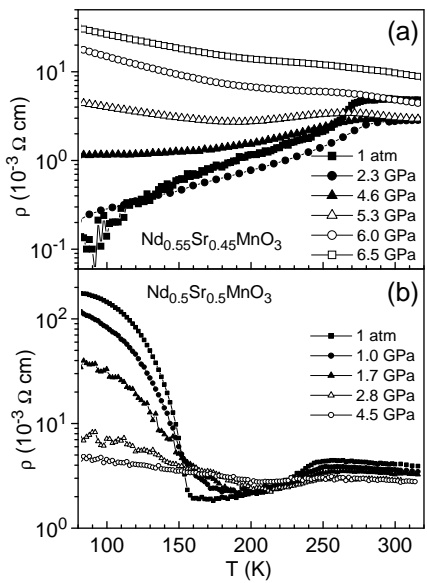}\\
\caption{\label{fig-1}Resistivity of (a) Nd$_{0.55}$Sr$_{0.45}$MnO$_{3}$ and (b) Nd$_{0.5}$Sr$_{0.5}$MnO$_{3}$ under pressure.}
\end{figure}
\begin{figure}
\includegraphics[width=2.2in]{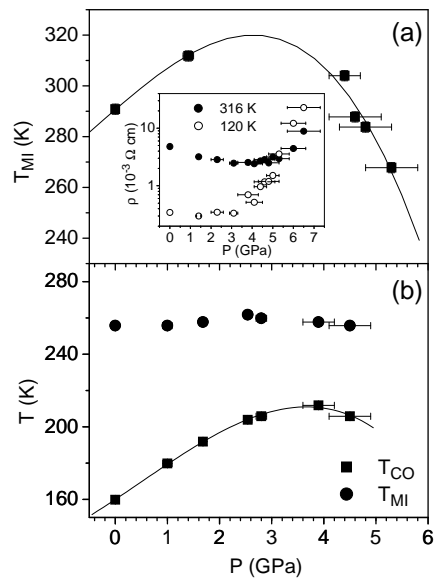}\\
\caption{\label{fig-2}Transition temperatures of (a) Nd$_{0.55}$Sr$_{0.45}$MnO$_{3}$ and (b) Nd$_{0.5}$Sr$_{0.5}$MnO$_{3}$. The solid lines are third-order polynomial fits as guides to the eyes. (a) Metal-insulator transition temperature of Nd$_{0.55}$Sr$_{0.45}$MnO$_{3}$ where the inset shows the resistivity changes with pressure in PMI phase [at 316 K (solid circle)] and FMM phase [at 120 K (open circle)]; (b) Metal-insulator transition (solid circle) and charge-ordering transition (solid square) temperatures of Nd$_{0.5}$Sr$_{0.5}$MnO$_{3}$.}
\end{figure}

Nd$_{0.55}$Sr$_{0.45}$MnO$_{3}$ is a double exchange compound, with a FMM to PMI transition at $\sim$280 K upon warming. Under pressure, the electron transport is modified in an interesting manner. Figure \ref{fig-1}(a) shows the resistivity as a function of temperature and pressure. The most important feature is the insulating state arising at low temperature under pressure. With pressure increase, the insulating behavior dominates above $\sim$6 GPa. Consequently, the resistivity in the measured temperature range changes with pressure [inset of Fig.\  \ref{fig-2}(a)]. Below $\sim$3.5 GPa, the resistivity in the PMI phase is reduced while in the FMM phase it is almost unchanged. Above $\sim$3.5 GPa, in both phases, resistivity increases rapidly with pressure. In the low-pressure range, the metal-insulator transition temperature T$_{MI}$ increases with pressure. Due to the limit of the instrument, T$_{MI}$ above 325 K in the range of $\sim$2-4 GPa cannot be determined. Above $\sim$4 GPa, the transition temperature decreases on pressure increase. Above $\sim$6 GPa, the insulating behavior dominates so that the transition temperature cannot be determined, although there is still a trace of metallic behavior. The change of transition temperature with pressure is plotted in Fig.\  \ref{fig-2}(a). The third-order polynomial fit gives a critical pressure P*$\sim$2.6 GPa, while the resistivity in paramagnetic phase (at $\sim$316 K) gives P*$\sim$3.6 GPa [inset of Fig.\  \ref{fig-2}(a)].

In this compound, the behavior of T$_{MI}$ and resistivity is similar to that found in La$_{0.60}$Y$_{0.07}$Ca$_{0.33}$MnO$_{3}$ (Ref.\ \onlinecite{cui_prb_67_104107_03}) and Pr$_{0.7}$Ca$_{0.3}$MnO$_{3}$.\cite{cui_apl_83_2856_03} However, the mechanism by which the materials become insulating at high-pressures is different. In those two compounds, the materials become insulating through the suppression of the FMM state, displaying a decreasing T$_{MI}$ above critical pressure. In Nd$_{0.55}$Sr$_{0.45}$MnO$_{3}$, the insulating state at high pressures has two origins: the suppression of the FMM state and the expansion of a low temperature insulating phase, which appears with pressure increase and finally dominates at high pressures. The enhancement of the insulating phase is the dominant contribution to the change in resistivity.

Abramovich \textit{et al}.\cite{abramovich_jpcm_12_627_00} proposed a phase-separation model in which the AFI droplets lie in a conducting ferromagnetic host. In the phase-separation model, the behavior of the material becoming insulating can be understood as pressure induced percolation where the increasing pressure suppresses the FMM component and enhances AFI component above P*.

It is noted that the transport behavior at high pressures where the compound becomes insulating is similar to that of Nd$_{0.45}$Sr$_{0.55}$MnO$_{3}$ at ambient pressure.\cite{kuwahara_prl_82_4316_99} When high magnetic field of 35 T is applied, the resistivity of the A-type antiferromagnetic metallic Nd$_{0.45}$Sr$_{0.55}$MnO$_{3}$ becomes similar to that of Nd$_{0.55}$Sr$_{0.45}$MnO$_{3}$, which is ascribed to the destruction of the A-type antiferromagnetic spin ordering and $d_{x^{2}-y^{2}}$ orbital ordering.\cite{hayashi_prb_65_024408_02} Considering the similarity between the resistivity of x = 0.45 compound under high-pressure and x = 0.55 at ambient pressure\cite{kuwahara_prl_82_4316_99} and that of x = 0.55 in high magnetic field,\cite{hayashi_prb_65_024408_02} one might speculate that the state induced by pressure in x = 0.45 compound has a similar spin and orbital structure to Nd$_{0.45}$Sr$_{0.55}$MnO$_{3}$.

\begin{figure}
\includegraphics[width=2.5in]{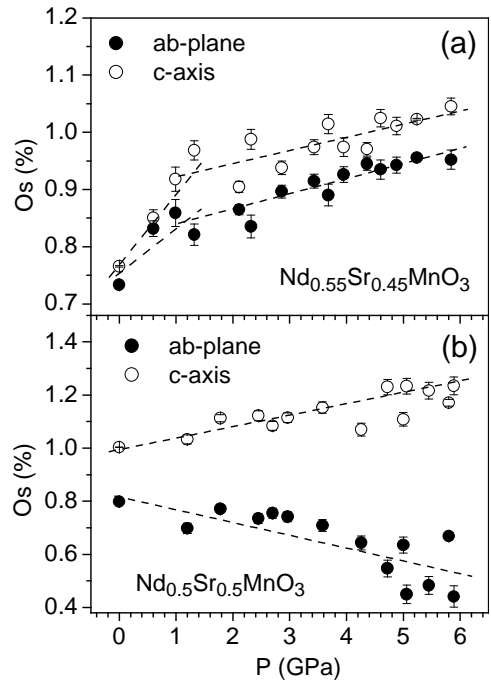}\\
\caption{\label{fig-3}Pressure dependence of the ab-plane and c-axis orthorhombic distortions of (a) Nd$_{0.55}$Sr$_{0.45}$MnO$_{3}$ and (b) Nd$_{0.5}$Sr$_{0.5}$MnO$_{3}$. (The dashed lines are guides to the eye.)}
\end{figure}

With the high-pressure structural measurements, it is found that the lattice is compressed anisotropically by pressure. The varying rates of change of lattice parameters under pressure lead to further distortion of the unit cell. To describe the orthorhombic distortion, Meneghini \textit{et al}.\cite{meneghini_prb_65_012111_02} defined the ab-plane distortion ($Os_{ab}=2\frac{a-b}{a+b}$) and c-axis distortion ($O_{c}=2\frac{a+b-c\sqrt{2}}{a+b+c\sqrt{2}}$) (in Pbnm symmetry). When the lattice is cubic, both Os$_{ab}$ and O$_{c}$ are zero. Figure \ref{fig-3}(a) shows that both distortions increase under pressure, indicating a more distorted structure from the cubic case.

The structure of Nd$_{0.45}$Sr$_{0.55}$MnO$_{3}$ is O$^{\ddag}$ ($a\approx b<\frac{c}{\sqrt2}$).\cite{kajimoto_prb_60_9506_99} The corresponding orthorhombic distortion is $\sim$0 in the ab-plane and -2\% along the c-axis (calculated with the data in Ref.\  \onlinecite{kajimoto_prb_60_9506_99}), which corresponds to the $d_{x^{2}-y^{2}}$-type orbital ordering and the A-type antiferromagnetic metal state. Under pressure, the orthorhombic distortion for both the c-axis and the ab-plane increase in Nd$_{0.55}$Sr$_{0.45}$MnO$_{3}$, indicating that the high-pressure structure is more different from Nd$_{0.45}$Sr$_{0.55}$MnO$_{3}$ than at ambient pressure. The ab-plane distortion increase of Nd$_{0.55}$Sr$_{0.45}$MnO$_{3}$ under pressure implies that the orbital state is different from that of Nd$_{0.45}$Sr$_{0.55}$MnO$_{3}$ at ambient pressure. However, the similarity between the resistivities (in both absolute value and shape) seems to suggest that pressure induces an A-type antiferromagnetic state in Nd$_{0.55}$Sr$_{0.45}$MnO$_{3}$.

\subsection{Nd$_{0.5}$Sr$_{0.5}$MnO$_{3}$}
Figure \ref{fig-1}(b) shows the resistivity of Nd$_{0.5}$Sr$_{0.5}$MnO$_{3}$ under pressure. In the low temperature CO AFI phase, the resistivity is reduced by pressure. On the other hand, the insulating region extends to higher temperature so that the ferromagnetic metallic state is suppressed. If the temperature where insulating and metallic states cross (the resistivity minimum) is defined as CO transition temperature, T$_{CO}$ increases first with pressure and appears to decrease above $\sim$3.8 GPa [Fig.\  \ref{fig-2}(b)]. At the same time, pressure affects the metal-insulator transition only slightly. With pressure increase, T$_{MI}$ increases below $\sim$3 GPa and drops above $\sim$3 GPa. The highest T$_{MI}$ is only $\sim$4 K higher than that at ambient pressure. In the measured pressure range, resistivity in the PMI phase is suppressed.

T$_{CO}$ increases with pressure below $\sim$3.8 GPa while T$_{MI}$ shows almost no change [Fig.\  \ref{fig-2}(b)]. This is different from the result that hydrostatic pressure ($<$1 GPa) increases T$_{C}$ and decreases T$_{CO}$ reported by other authors in single crystals.\cite{moritomo_jpsj_66_556_97} On the contrary, our results are consistent with the effects of uniaxial pressure along the c-axis.\cite{arima_prb_60_15013_99} In Fig.\  \ref{fig-3}(b), it is seen that the c-axis distortion increases while the ab-plane distortion decreases with pressure. Because the CO state corresponds to a higher orthorhombic distortion state,\cite{kuwahara_sci_270_961_95} possibly the pressure induced increase of c-axis distortion enhances the CO state. On the other hand, the decrease in the ab-plane distortion may enhance the electron hoping and lead to the resistivity decrease in the ab-plane.

Roy \textit{et al}.\cite{roy_prb_63_094416_01} reported that pressure above $\sim$1.5 GPa splits the coincident antiferromagnetic and charge-ordering transitions in which T$_{CO}$ increases while T$_{N}$ decreases. With the transitions decoupled, resistivity rises abruptly at the magnetic transition but not at the CO transition, implying that low temperature resistivity comes mostly from the CE-type antiferromagnetic state. We did not observe the T$_{CO}$ and T$_{N}$ splitting in a larger pressure range, possibility because our sample is polycrystalline. In this case, the grain size distribution and randomly distributed grain orientations may lead to the broader CO AFI transition in polycrystalline samples\cite{caignaert_solidstate_99_173_96} than in single crystal.\cite{kuwahara_sci_270_961_95} However, the large suppression of resistivity indicates that the antiferromagnetic state, specifically the CE-type antiferromagnetic state, is suppressed, which is also consistent with the ab-plane orthorhombic distortion reduction [Fig.\  \ref{fig-3}(b)].

In Nd$_{0.5}$Sr$_{0.5}$MnO$_{3}$, by substituting Nd$^{3+}$ with larger size La$^{3+}$, the bandwidth is increased and hence the CO phase is suppressed and T$_{C}$ increases. With applied pressure, a transition from the CO CE-type AFI to the A-type AFI was suggested, in which resistivity is suppressed and T$_{CO}$ gradually increases with pressure.\cite{moritomo_prb_55_7549_97} This is consistent with our result in the parent compound Nd$_{0.5}$Sr$_{0.5}$MnO$_{3}$ and at a much higher pressure. The smaller ab-plane distortion and larger c-axis distortion at high pressures may favor an A-type antiferromagnetic state and $d_{x^{2}-y^{2}}$ orbital ordering as in the x = 0.55 compound. In the A-type AFI state, resistivity is decreased due to enhanced in-plane transfer integral by the reduction of in-plane distortion. In the phase-separation model,\cite{ritter_prb_61_9229_00} the A-type antiferromagnetic phase is enhanced and the CO CE-type antiferromagnetic phase is suppressed concomitantly by pressure. Because the bandwidth is sensitive to the atomic structure of the MnO$_{6}$ octahedra, especially the Mn-O-Mn bond angle, it is highly desired to measure the atomic structure to explain the electronic and magnetic behavior under pressure. Unfortunately, it has been found difficult to acquire the atomic structure information from x-ray diffraction, due to the large difference between the scattering factors of the Nd/Sr and oxygen atoms compared to the case of La$_{0.6}$Y$_{0.07}$Ca$_{0.33}$MnO$_{3}$ in our previous work,\cite{cui_prb_67_104107_03} which makes the refinement to the oxygen coordinates in the unit cell extremely difficult. A similar case also exists in the study on the structure of PbO under high-pressure.\cite{haussermann_acie_40_4624_01} So other experimental techniques, such as high-pressure Raman scattering, are proposed to probe the local structure changes under pressure.

\begin{figure}
\includegraphics[width=2.8in]{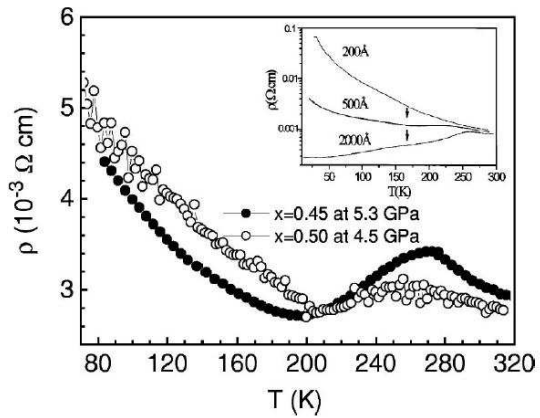}\\
\caption{\label{fig-4}Comparison of resistivity of Nd$_{0.55}$Sr$_{0.45}$MnO$_{3}$ and Nd$_{0.5}$Sr$_{0.5}$MnO$_{3}$ under pressure. The inset shows the resistivity of Nd$_{0.5}$Sr$_{0.5}$MnO$_{3}$ thin films with different thicknesses from Ref.\  \onlinecite{prellier_apl_75_397_99} for comparison with our pressure results.}
\end{figure}

The two Nd$_{1-x}$Sr$_{x}$MnO$_{3}$ manganites at x = 0.45 and 0.5 have very different electronic, magnetic, and orbital ground states at ambient conditions. However, when we compare the resistivity at high-pressure, we can find a surprising similarity between them [Figs.\  \ref{fig-1}(a) and \ref{fig-1}(b)]. Figure \ref{fig-4} is an example of the resistivity of these two compounds at pressures above the critical pressure. The similarity also seems to imply a similar electronic and magnetic state. The structural measurements partly justify this assumption. The orthorhombic distortion of x = 0.45 is increased by pressure to almost the same as that of the x = 0.5 compound at ambient pressure [Figs.\  \ref{fig-3}(a) and \ref{fig-3}(b)]. In addition, the similarity between the high-pressure resistivity of these two compounds and that of Nd$_{0.45}$Sr$_{0.55}$MnO$_{3}$ also suggests an A-type AFI phase in the high-pressure phase.

It is interesting to compare the effects of pressure and strain in thin films. The inset of Fig.\  \ref{fig-4} shows the resistivity of Nd$_{0.5}$Sr$_{0.5}$MnO$_{3}$ thin films of several typical thicknesses from Prellier \textit{et al}.\cite{prellier_apl_75_397_99} With thickness decrease, the strain in thin films is found to increase.\cite{qian_prb_63_224424_01} Compared with Nd$_{0.55}$Sr$_{0.45}$MnO$_{3}$, the resistivity evolution with thickness decrease (strain increase) is in analogy to pressure increase in bulk Nd$_{0.55}$Sr$_{0.45}$MnO$_{3}$, indicating that pressure increases strain in bulk sample proved by structure measurements.

\section{Summary}
In summary, by studying the resistivity and structure of Nd$_{1-x}$Sr$_{x}$MnO$_{3}$ (x = 0.45, 0.5) at high pressures, it is found that they have a similar resistivity as a function of temperature, which results from the different effects of pressure on their structures. Under pressure, both the ferromagnetic metallic state in the x = 0.45 compound and the CE-type antiferromagnetic insulating state in the x = 0.5 compound are suppressed. By comparing the resistivity and structure with the x = 0.55 compound, pressure appears to induce a similar electronic and magnetic state in these two compounds with much different ground states. We suggest that the pressure induced magnetic states in both samples are A-type antiferromagnetic.

\begin{acknowledgments}
The high-pressure x-ray diffraction measurements were performed at beamline X17B1, NSLS, Brookhaven National Laboratory which is supported by the U.S.\ Department of Energy under Contract No.\ DE-AC02-98CH10886. The authors would like to thank Dr. Jingzhu Hu at X17C, NSLS for her assistance on the pressure calibration for x-ray diffraction. This work is supported by the National Science Foundation under Grant No.\ DMR-0209243.
\end{acknowledgments}

\bibliography{BE9062-cui}

\begin{thebibliography}{26}
\expandafter\ifx\csname natexlab\endcsname\relax\def\natexlab#1{#1}\fi
\expandafter\ifx\csname bibnamefont\endcsname\relax
  \def\bibnamefont#1{#1}\fi
\expandafter\ifx\csname bibfnamefont\endcsname\relax
  \def\bibfnamefont#1{#1}\fi
\expandafter\ifx\csname citenamefont\endcsname\relax
  \def\citenamefont#1{#1}\fi
\expandafter\ifx\csname url\endcsname\relax
  \def\url#1{\texttt{#1}}\fi
\expandafter\ifx\csname urlprefix\endcsname\relax\def\urlprefix{URL }\fi
\providecommand{\bibinfo}[2]{#2}
\providecommand{\eprint}[2][]{\url{#2}}

\bibitem[{\citenamefont{Tokura and Tomioka}(1999)}]{tokura_jmmm_200_1_99}
\bibinfo{author}{\bibfnamefont{Y.}~\bibnamefont{Tokura}} \bibnamefont{and}
  \bibinfo{author}{\bibfnamefont{Y.}~\bibnamefont{Tomioka}},
  \bibinfo{journal}{J. Magn. Magn. Mater.} \textbf{\bibinfo{volume}{200}},
  \bibinfo{pages}{1} (\bibinfo{year}{1999}).

\bibitem[{\citenamefont{Kawano et~al.}(1997)\citenamefont{Kawano, Kajimoto,
  Yoshizawa, Tomioka, Kuwahara, and Tokura}}]{kawano_prl_78_4253_97}
\bibinfo{author}{\bibfnamefont{H.}~\bibnamefont{Kawano}},
  \bibinfo{author}{\bibfnamefont{R.}~\bibnamefont{Kajimoto}},
  \bibinfo{author}{\bibfnamefont{H.}~\bibnamefont{Yoshizawa}},
  \bibinfo{author}{\bibfnamefont{Y.}~\bibnamefont{Tomioka}},
  \bibinfo{author}{\bibfnamefont{H.}~\bibnamefont{Kuwahara}}, \bibnamefont{and}
  \bibinfo{author}{\bibfnamefont{Y.}~\bibnamefont{Tokura}},
  \bibinfo{journal}{Phys. Rev. Lett.} \textbf{\bibinfo{volume}{78,}},
  \bibinfo{pages}{4253} (\bibinfo{year}{1997}).

\bibitem[{\citenamefont{Kuwahara et~al.}(1995)\citenamefont{Kuwahara, Tomioka,
  Asamitsu, Moritomo, and Tokura}}]{kuwahara_sci_270_961_95}
\bibinfo{author}{\bibfnamefont{H.}~\bibnamefont{Kuwahara}},
  \bibinfo{author}{\bibfnamefont{Y.}~\bibnamefont{Tomioka}},
  \bibinfo{author}{\bibfnamefont{A.}~\bibnamefont{Asamitsu}},
  \bibinfo{author}{\bibfnamefont{Y.}~\bibnamefont{Moritomo}}, \bibnamefont{and}
  \bibinfo{author}{\bibfnamefont{Y.}~\bibnamefont{Tokura}},
  \bibinfo{journal}{Science} \textbf{\bibinfo{volume}{270}},
  \bibinfo{pages}{961} (\bibinfo{year}{1995}).

\bibitem[{\citenamefont{Mahendiran et~al.}(1999)\citenamefont{Mahendiran,
  Ibarra, Maignan, Millange, Arulraj, Mahesh, Raveau, and
  Rao}}]{mahendiran_prl_82_2191_99}
\bibinfo{author}{\bibfnamefont{R.}~\bibnamefont{Mahendiran}},
  \bibinfo{author}{\bibfnamefont{M.~R.} \bibnamefont{Ibarra}},
  \bibinfo{author}{\bibfnamefont{A.}~\bibnamefont{Maignan}},
  \bibinfo{author}{\bibfnamefont{F.}~\bibnamefont{Millange}},
  \bibinfo{author}{\bibfnamefont{A.}~\bibnamefont{Arulraj}},
  \bibinfo{author}{\bibfnamefont{R.}~\bibnamefont{Mahesh}},
  \bibinfo{author}{\bibfnamefont{B.}~\bibnamefont{Raveau}}, \bibnamefont{and}
  \bibinfo{author}{\bibfnamefont{C.~N.~R.} \bibnamefont{Rao}},
  \bibinfo{journal}{Phys. Rev. Lett.} \textbf{\bibinfo{volume}{82}},
  \bibinfo{pages}{2191} (\bibinfo{year}{1999}).

\bibitem[{\citenamefont{Nakamura et~al.}(1999)\citenamefont{Nakamura, Arima,
  Nakazawa, Wakabayashi, and Murakami}}]{nakamura_prb_60_2425_99}
\bibinfo{author}{\bibfnamefont{K.}~\bibnamefont{Nakamura}},
  \bibinfo{author}{\bibfnamefont{T.}~\bibnamefont{Arima}},
  \bibinfo{author}{\bibfnamefont{A.}~\bibnamefont{Nakazawa}},
  \bibinfo{author}{\bibfnamefont{Y.}~\bibnamefont{Wakabayashi}},
  \bibnamefont{and} \bibinfo{author}{\bibfnamefont{Y.}~\bibnamefont{Murakami}},
  \bibinfo{journal}{Phys. Rev. B} \textbf{\bibinfo{volume}{60}},
  \bibinfo{pages}{2425} (\bibinfo{year}{1999}).

\bibitem[{\citenamefont{Zvyagin et~al.}(2002)\citenamefont{Zvyagin, Angerhofer,
  Kamenev, Brunel, Balakrishnan, and Paul}}]{zayagin_solidstate_121_117_02}
\bibinfo{author}{\bibfnamefont{S.}~\bibnamefont{Zvyagin}},
  \bibinfo{author}{\bibfnamefont{A.}~\bibnamefont{Angerhofer}},
  \bibinfo{author}{\bibfnamefont{K.~V.} \bibnamefont{Kamenev}},
  \bibinfo{author}{\bibfnamefont{L.-C.} \bibnamefont{Brunel}},
  \bibinfo{author}{\bibfnamefont{G.}~\bibnamefont{Balakrishnan}},
  \bibnamefont{and} \bibinfo{author}{\bibfnamefont{D.~M.} \bibnamefont{Paul}},
  \bibinfo{journal}{Solid State Commun.} \textbf{\bibinfo{volume}{121}},
  \bibinfo{pages}{117} (\bibinfo{year}{2002}).

\bibitem[{\citenamefont{Kuwahara et~al.}(1998)\citenamefont{Kuwahara, Okuda,
  Tomioka, Kimura, Asamitsu, and Tokura}}]{kuwahara_mrssp_494_83_98}
\bibinfo{author}{\bibfnamefont{H.}~\bibnamefont{Kuwahara}},
  \bibinfo{author}{\bibfnamefont{T.}~\bibnamefont{Okuda}},
  \bibinfo{author}{\bibfnamefont{Y.}~\bibnamefont{Tomioka}},
  \bibinfo{author}{\bibfnamefont{T.}~\bibnamefont{Kimura}},
  \bibinfo{author}{\bibfnamefont{A.}~\bibnamefont{Asamitsu}}, \bibnamefont{and}
  \bibinfo{author}{\bibfnamefont{Y.}~\bibnamefont{Tokura}},
  \bibinfo{journal}{Mat. Res. Soc. Symp. Proc.} \textbf{\bibinfo{volume}{494}},
  \bibinfo{pages}{83} (\bibinfo{year}{1998}).

\bibitem[{\citenamefont{Hayashi et~al.}(2002)\citenamefont{Hayashi, Miura,
  Noda, Kuwahara, Okamoto, Ishihara, and Maekawa}}]{hayashi_prb_65_024408_02}
\bibinfo{author}{\bibfnamefont{T.}~\bibnamefont{Hayashi}},
  \bibinfo{author}{\bibfnamefont{N.}~\bibnamefont{Miura}},
  \bibinfo{author}{\bibfnamefont{K.}~\bibnamefont{Noda}},
  \bibinfo{author}{\bibfnamefont{H.}~\bibnamefont{Kuwahara}},
  \bibinfo{author}{\bibfnamefont{S.}~\bibnamefont{Okamoto}},
  \bibinfo{author}{\bibfnamefont{S.}~\bibnamefont{Ishihara}}, \bibnamefont{and}
  \bibinfo{author}{\bibfnamefont{S.}~\bibnamefont{Maekawa}},
  \bibinfo{journal}{Phys. Rev. B} \textbf{\bibinfo{volume}{65}},
  \bibinfo{pages}{024408} (\bibinfo{year}{2002}).

\bibitem[{\citenamefont{Eremenko et~al.}(2001)\citenamefont{Eremenko,
  Gnatchenko, Makedonska, Shabakayeva, Shvedun, Sirenko, Fink-Finowicki,
  Kamenev, Balakrishnan, and Paul}}]{eremenko_ltp_27_930_01}
\bibinfo{author}{\bibfnamefont{V.}~\bibnamefont{Eremenko}},
  \bibinfo{author}{\bibfnamefont{S.}~\bibnamefont{Gnatchenko}},
  \bibinfo{author}{\bibfnamefont{N.}~\bibnamefont{Makedonska}},
  \bibinfo{author}{\bibfnamefont{Y.}~\bibnamefont{Shabakayeva}},
  \bibinfo{author}{\bibfnamefont{M.}~\bibnamefont{Shvedun}},
  \bibinfo{author}{\bibfnamefont{V.}~\bibnamefont{Sirenko}},
  \bibinfo{author}{\bibfnamefont{J.}~\bibnamefont{Fink-Finowicki}},
  \bibinfo{author}{\bibfnamefont{K.~V.} \bibnamefont{Kamenev}},
  \bibinfo{author}{\bibfnamefont{G.}~\bibnamefont{Balakrishnan}},
  \bibnamefont{and} \bibinfo{author}{\bibfnamefont{D.~M.} \bibnamefont{Paul}},
  \bibinfo{journal}{Low Temp. Phys.} \textbf{\bibinfo{volume}{27}},
  \bibinfo{pages}{930} (\bibinfo{year}{2001}).

\bibitem[{\citenamefont{Laffez et~al.}(1996)\citenamefont{Laffez, Tendeloo,
  Millange, Caignaert, Hervieu, and Raveau}}]{laffez_mrb_31_905_96}
\bibinfo{author}{\bibfnamefont{P.}~\bibnamefont{Laffez}},
  \bibinfo{author}{\bibfnamefont{G.~V.} \bibnamefont{Tendeloo}},
  \bibinfo{author}{\bibfnamefont{F.}~\bibnamefont{Millange}},
  \bibinfo{author}{\bibfnamefont{V.}~\bibnamefont{Caignaert}},
  \bibinfo{author}{\bibfnamefont{M.}~\bibnamefont{Hervieu}}, \bibnamefont{and}
  \bibinfo{author}{\bibfnamefont{B.}~\bibnamefont{Raveau}},
  \bibinfo{journal}{Mater. Res. Bull.} \textbf{\bibinfo{volume}{31}},
  \bibinfo{pages}{905} (\bibinfo{year}{1996}).

\bibitem[{\citenamefont{Ritter et~al.}(2000)\citenamefont{Ritter, Mahendiran,
  Ibarra, Morellon, Maignan, Raveau, and Rao}}]{ritter_prb_61_9229_00}
\bibinfo{author}{\bibfnamefont{C.}~\bibnamefont{Ritter}},
  \bibinfo{author}{\bibfnamefont{R.}~\bibnamefont{Mahendiran}},
  \bibinfo{author}{\bibfnamefont{M.~R.} \bibnamefont{Ibarra}},
  \bibinfo{author}{\bibfnamefont{L.}~\bibnamefont{Morellon}},
  \bibinfo{author}{\bibfnamefont{A.}~\bibnamefont{Maignan}},
  \bibinfo{author}{\bibfnamefont{B.}~\bibnamefont{Raveau}}, \bibnamefont{and}
  \bibinfo{author}{\bibfnamefont{C.~N.~R.} \bibnamefont{Rao}},
  \bibinfo{journal}{Phys. Rev. B} \textbf{\bibinfo{volume}{61}},
  \bibinfo{pages}{R9229} (\bibinfo{year}{2000}).

\bibitem[{\citenamefont{Kajimoto et~al.}(1999)\citenamefont{Kajimoto,
  Yoshizawa, Kawano, Kuwahara, Tokura, Ohoyama, and
  Ohashi}}]{kajimoto_prb_60_9506_99}
\bibinfo{author}{\bibfnamefont{R.}~\bibnamefont{Kajimoto}},
  \bibinfo{author}{\bibfnamefont{H.}~\bibnamefont{Yoshizawa}},
  \bibinfo{author}{\bibfnamefont{H.}~\bibnamefont{Kawano}},
  \bibinfo{author}{\bibfnamefont{H.}~\bibnamefont{Kuwahara}},
  \bibinfo{author}{\bibfnamefont{Y.}~\bibnamefont{Tokura}},
  \bibinfo{author}{\bibfnamefont{K.}~\bibnamefont{Ohoyama}}, \bibnamefont{and}
  \bibinfo{author}{\bibfnamefont{M.}~\bibnamefont{Ohashi}},
  \bibinfo{journal}{Phys. Rev. B} \textbf{\bibinfo{volume}{60}},
  \bibinfo{pages}{9506} (\bibinfo{year}{1999}).

\bibitem[{\citenamefont{Moritomo
  et~al.}(1997{\natexlab{a}})\citenamefont{Moritomo, Kuwahara, and
  Tokura}}]{moritomo_jpsj_66_556_97}
\bibinfo{author}{\bibfnamefont{Y.}~\bibnamefont{Moritomo}},
  \bibinfo{author}{\bibfnamefont{H.}~\bibnamefont{Kuwahara}}, \bibnamefont{and}
  \bibinfo{author}{\bibfnamefont{Y.}~\bibnamefont{Tokura}},
  \bibinfo{journal}{J. Phys. Soc. Jpn.} \textbf{\bibinfo{volume}{66}},
  \bibinfo{pages}{556} (\bibinfo{year}{1997}{\natexlab{a}}).

\bibitem[{\citenamefont{hisa Arima and Nakamura}(1999)}]{arima_prb_60_15013_99}
\bibinfo{author}{\bibfnamefont{T.}~\bibnamefont{hisa Arima}} \bibnamefont{and}
  \bibinfo{author}{\bibfnamefont{K.}~\bibnamefont{Nakamura}},
  \bibinfo{journal}{Phys. Rev. B} \textbf{\bibinfo{volume}{60}},
  \bibinfo{pages}{R15013} (\bibinfo{year}{1999}).

\bibitem[{\citenamefont{Prellier et~al.}(1999)\citenamefont{Prellier, Biswas,
  Rajeswari, Venkatesan, and Greene}}]{prellier_apl_75_397_99}
\bibinfo{author}{\bibfnamefont{W.}~\bibnamefont{Prellier}},
  \bibinfo{author}{\bibfnamefont{A.}~\bibnamefont{Biswas}},
  \bibinfo{author}{\bibfnamefont{M.}~\bibnamefont{Rajeswari}},
  \bibinfo{author}{\bibfnamefont{T.}~\bibnamefont{Venkatesan}},
  \bibnamefont{and} \bibinfo{author}{\bibfnamefont{R.~L.}
  \bibnamefont{Greene}}, \bibinfo{journal}{Appl. Phys. Lett.}
  \textbf{\bibinfo{volume}{75}}, \bibinfo{pages}{397} (\bibinfo{year}{1999}).

\bibitem[{\citenamefont{Qian et~al.}(2001)\citenamefont{Qian, Tyson, Kao,
  Prellier, Bai, Biswas, and Greene}}]{qian_prb_63_224424_01}
\bibinfo{author}{\bibfnamefont{Q.}~\bibnamefont{Qian}},
  \bibinfo{author}{\bibfnamefont{T.~A.} \bibnamefont{Tyson}},
  \bibinfo{author}{\bibfnamefont{C.-C.} \bibnamefont{Kao}},
  \bibinfo{author}{\bibfnamefont{W.}~\bibnamefont{Prellier}},
  \bibinfo{author}{\bibfnamefont{J.}~\bibnamefont{Bai}},
  \bibinfo{author}{\bibfnamefont{A.}~\bibnamefont{Biswas}}, \bibnamefont{and}
  \bibinfo{author}{\bibfnamefont{R.~L.} \bibnamefont{Greene}},
  \bibinfo{journal}{Phys. Rev. B} \textbf{\bibinfo{volume}{63}},
  \bibinfo{pages}{224424} (\bibinfo{year}{2001}).

\bibitem[{\citenamefont{Caignaert et~al.}(1996)\citenamefont{Caignaert,
  Millange, Hervieu, Suard, and Raveau}}]{caignaert_solidstate_99_173_96}
\bibinfo{author}{\bibfnamefont{V.}~\bibnamefont{Caignaert}},
  \bibinfo{author}{\bibfnamefont{F.}~\bibnamefont{Millange}},
  \bibinfo{author}{\bibfnamefont{M.}~\bibnamefont{Hervieu}},
  \bibinfo{author}{\bibfnamefont{E.}~\bibnamefont{Suard}}, \bibnamefont{and}
  \bibinfo{author}{\bibfnamefont{B.}~\bibnamefont{Raveau}},
  \bibinfo{journal}{Solid State Commun.} \textbf{\bibinfo{volume}{99}},
  \bibinfo{pages}{173} (\bibinfo{year}{1996}).

\bibitem[{\citenamefont{Tomioka et~al.}(1997)\citenamefont{Tomioka, Kuwahara,
  Asamitsu, Kasai, and Tokura}}]{tomioka_prl_70_3609_97}
\bibinfo{author}{\bibfnamefont{Y.}~\bibnamefont{Tomioka}},
  \bibinfo{author}{\bibfnamefont{H.}~\bibnamefont{Kuwahara}},
  \bibinfo{author}{\bibfnamefont{A.}~\bibnamefont{Asamitsu}},
  \bibinfo{author}{\bibfnamefont{M.}~\bibnamefont{Kasai}}, \bibnamefont{and}
  \bibinfo{author}{\bibfnamefont{Y.}~\bibnamefont{Tokura}},
  \bibinfo{journal}{Appl. Phys. Lett.} \textbf{\bibinfo{volume}{70}},
  \bibinfo{pages}{3609} (\bibinfo{year}{1997}).

\bibitem[{\citenamefont{Cui et~al.}(2003)\citenamefont{Cui, Tyson, Zhong,
  Carlo, and Qin}}]{cui_prb_67_104107_03}
\bibinfo{author}{\bibfnamefont{C.}~\bibnamefont{Cui}},
  \bibinfo{author}{\bibfnamefont{T.~A.} \bibnamefont{Tyson}},
  \bibinfo{author}{\bibfnamefont{Z.}~\bibnamefont{Zhong}},
  \bibinfo{author}{\bibfnamefont{J.~P.} \bibnamefont{Carlo}}, \bibnamefont{and}
  \bibinfo{author}{\bibfnamefont{Y.}~\bibnamefont{Qin}},
  \bibinfo{journal}{Phys. Rev. B} \textbf{\bibinfo{volume}{67}},
  \bibinfo{pages}{104107} (\bibinfo{year}{2003}).

\bibitem[{\citenamefont{Cui and Tyson}(2003)}]{cui_apl_83_2856_03}
\bibinfo{author}{\bibfnamefont{C.}~\bibnamefont{Cui}} \bibnamefont{and}
  \bibinfo{author}{\bibfnamefont{T.~A.} \bibnamefont{Tyson}},
  \bibinfo{journal}{Appl. Phys. Lett.} \textbf{\bibinfo{volume}{83}},
  \bibinfo{pages}{2856} (\bibinfo{year}{2003}).

\bibitem[{\citenamefont{Abramovich et~al.}(2000)\citenamefont{Abramovich,
  Michurin, Gorbenko, and Kaul}}]{abramovich_jpcm_12_627_00}
\bibinfo{author}{\bibfnamefont{A.~I.} \bibnamefont{Abramovich}},
  \bibinfo{author}{\bibfnamefont{A.~V.} \bibnamefont{Michurin}},
  \bibinfo{author}{\bibfnamefont{O.~Y.} \bibnamefont{Gorbenko}},
  \bibnamefont{and} \bibinfo{author}{\bibfnamefont{A.~R.} \bibnamefont{Kaul}},
  \bibinfo{journal}{J. Phys.: Condens. Matter} \textbf{\bibinfo{volume}{12}},
  \bibinfo{pages}{L627} (\bibinfo{year}{2000}).

\bibitem[{\citenamefont{Kuwahara et~al.}(1999)\citenamefont{Kuwahara, Okuda,
  Tomioka, Asamitsu, and Tokura}}]{kuwahara_prl_82_4316_99}
\bibinfo{author}{\bibfnamefont{H.}~\bibnamefont{Kuwahara}},
  \bibinfo{author}{\bibfnamefont{T.}~\bibnamefont{Okuda}},
  \bibinfo{author}{\bibfnamefont{Y.}~\bibnamefont{Tomioka}},
  \bibinfo{author}{\bibfnamefont{A.}~\bibnamefont{Asamitsu}}, \bibnamefont{and}
  \bibinfo{author}{\bibfnamefont{Y.}~\bibnamefont{Tokura}},
  \bibinfo{journal}{Phys. Rev. Lett.} \textbf{\bibinfo{volume}{82}},
  \bibinfo{pages}{4316} (\bibinfo{year}{1999}).

\bibitem[{\citenamefont{Meneghini et~al.}(2002)\citenamefont{Meneghini, Levy,
  Mobilio, Ortolani, Nu{\~n}ez-Reguero, Kumar, and
  Sarma}}]{meneghini_prb_65_012111_02}
\bibinfo{author}{\bibfnamefont{C.}~\bibnamefont{Meneghini}},
  \bibinfo{author}{\bibfnamefont{D.}~\bibnamefont{Levy}},
  \bibinfo{author}{\bibfnamefont{S.}~\bibnamefont{Mobilio}},
  \bibinfo{author}{\bibfnamefont{M.}~\bibnamefont{Ortolani}},
  \bibinfo{author}{\bibfnamefont{M.}~\bibnamefont{Nu{\~n}ez-Reguero}},
  \bibinfo{author}{\bibfnamefont{A.}~\bibnamefont{Kumar}}, \bibnamefont{and}
  \bibinfo{author}{\bibfnamefont{D.~D.} \bibnamefont{Sarma}},
  \bibinfo{journal}{Phys. Rev. B} \textbf{\bibinfo{volume}{65}},
  \bibinfo{pages}{012111} (\bibinfo{year}{2002}).

\bibitem[{\citenamefont{Roy et~al.}(2001)\citenamefont{Roy, Husmann, Rosenbaum,
  and Mitchell}}]{roy_prb_63_094416_01}
\bibinfo{author}{\bibfnamefont{A.~S.} \bibnamefont{Roy}},
  \bibinfo{author}{\bibfnamefont{A.}~\bibnamefont{Husmann}},
  \bibinfo{author}{\bibfnamefont{T.~F.} \bibnamefont{Rosenbaum}},
  \bibnamefont{and} \bibinfo{author}{\bibfnamefont{J.~F.}
  \bibnamefont{Mitchell}}, \bibinfo{journal}{Phys. Rev. B}
  \textbf{\bibinfo{volume}{63}}, \bibinfo{pages}{094416}
  (\bibinfo{year}{2001}).

\bibitem[{\citenamefont{Moritomo
  et~al.}(1997{\natexlab{b}})\citenamefont{Moritomo, Kuwahara, Tomika, and
  Tokura}}]{moritomo_prb_55_7549_97}
\bibinfo{author}{\bibfnamefont{Y.}~\bibnamefont{Moritomo}},
  \bibinfo{author}{\bibfnamefont{H.}~\bibnamefont{Kuwahara}},
  \bibinfo{author}{\bibfnamefont{Y.}~\bibnamefont{Tomika}}, \bibnamefont{and}
  \bibinfo{author}{\bibfnamefont{Y.}~\bibnamefont{Tokura}},
  \bibinfo{journal}{Phys. Rev. B} \textbf{\bibinfo{volume}{55}},
  \bibinfo{pages}{7549} (\bibinfo{year}{1997}{\natexlab{b}}).

\bibitem[{\citenamefont{H{\"a}ussermann
  et~al.}(2001)\citenamefont{H{\"a}ussermann, Berastegui, Carlson, Haines, and
  L{\'e}ger}}]{haussermann_acie_40_4624_01}
\bibinfo{author}{\bibfnamefont{U.}~\bibnamefont{H{\"a}ussermann}},
  \bibinfo{author}{\bibfnamefont{P.}~\bibnamefont{Berastegui}},
  \bibinfo{author}{\bibfnamefont{S.}~\bibnamefont{Carlson}},
  \bibinfo{author}{\bibfnamefont{J.}~\bibnamefont{Haines}}, \bibnamefont{and}
  \bibinfo{author}{\bibfnamefont{J.-M.} \bibnamefont{L{\'e}ger}},
  \bibinfo{journal}{Angew. Chem. Int. Ed.} \textbf{\bibinfo{volume}{40}},
  \bibinfo{pages}{4624} (\bibinfo{year}{2001}).

\end{thebibliography}

\end{document}